\documentclass[%
 aip,
 amsmath,amssymb,
 reprint,%
]{revtex4-2}

\usepackage{graphicx}
\usepackage{dcolumn}
\usepackage{bm}

\usepackage[utf8]{inputenc}
\usepackage[T1]{fontenc}
\usepackage{mathptmx}
\usepackage{etoolbox}

\makeatletter
\def\@email#1#2{%
 \endgroup
 \patchcmd{\titleblock@produce}
  {\frontmatter@RRAPformat}
  {\frontmatter@RRAPformat{\produce@RRAP{*#1\href{mailto:#2}{#2}}}\frontmatter@RRAPformat}
  {}{}
}%
\makeatother
\begin{document}

\preprint{AIP/123-QED}

\title[Sample title]{Suppression and Regulation of Thermal Birefringence in Optical Voltage Sensor with Isomerism Electrodes and Arbitrary Electric Field Direction Modulation}
\author{Jun Li}

\author{Qifeng Xu}%

\affiliation{ 
	College of Electrical Engineering and Automation, Fuzhou University, Fuzhou, China
}

\author{Yifan Lin}

\affiliation{ 
	College of Electrical Engineering and Automation, Fuzhou University, Fuzhou, China
}

\author{Nan Xie*}

\email{t13056@fzu.edu.cn}
\affiliation{ 
	College of Electrical Engineering and Automation, Fuzhou University, Fuzhou, China
}%

\date{\today}

\begin{abstract}
The insufficient stability and reliability of Optical Voltage Sensor is primarily caused by thermal stress induced birefringence. In this paper, a method based on arbitrary electric field direction modulation and isomerism electrodes is proposed to  suppress or regulate it. 
With the aid of multi-physics Finite Element Method, Jones Matrix and the theory of photoelastic effect, it is found that metal or transparent isomerism electrodes can generate a special thermal stress distribution, which regulates the birefringence in the optical path and their induced measurement error.
The experiment is conducted on a 10mm cubic bismuth germanite  crystal, with cutting directions 110, $\bar{1}10$ and 001.
The experiment result shows that Cu isomerism electrodes with electric field angle of 59.9 degrees could generate 37\% less birefringence error  compared to parallel plate electrodes, in the temperature range from 25 $^{\circ} \text{C}$ to 40 $^{\circ} \text{C}$. However, the Indium Tin Oxide  electrodes with field angle of 29.6$^{\circ}$ produces  approximately 7 times error because of its bad ductility and thermal conduction. The proposed modeling and suppression method for birefringence is beneficial to design of high accuracy optical voltage sensor or electro-optical modulator.   
\end{abstract}

\maketitle

\section{\label{sec:level1}Introduction}
The Optical Voltage Sensor (OVS)$^{1-11}$ based on the Pockels effect offers advantages such as higher frequency response, wider dynamic range, excellent insulation performance, compact size, low insulation cost, light weight, and so on. However, its bad long-term stability and reliability have hindered its practical implementation. The main cause is the thermal stress induced birefringence$^{4, 12-16}$ in electro-optic crystals such as lithium niobate (LN) or bismuth germanite (BGO), as well as in the transmitting optical fibers. 

To effectively suppress the thermal stress birefringence, various methods have been proposed in recent years. The dual-path compensation method$^{16}$ utilizes a polarizing beam splitter  to separate the output light  into two orthogonal linearly polarized beams. The modulation depth of each beam exhibits an inverse relationship with temperature variation. By extracting the DC and AC components of both signals and performing signal processing, the DC component containing signals from stress birefringence could be eliminated. The dual-crystal method$^{17-19}$ involves serially connecting two nearly identical crystals in a specific manner. This arrangement reverses the fast and slow axes of the crystals, ensuring that the birefringence variation in the two crystals is opposite to each other, thus compensating the birefringence error in double electro-optical crystals. 

Another birefringence-suppression idea employs the image demodulation mode$^{20-23}$, which detects the rotation or displacement of images generated by the polarized light. To convert from phase delay to optical images, the precise and expensive optical elements such as crystal wedges$^{20}$, radial polarizer$^{21,22}$,  or S-wave plate$^{23}$ are needed. These modes usually have large linear phase measurement range, which  exceed to 180 degrees. Thus the method could enhance the intensity of effective signal and signal-noise ratio (SNR), reducing the affecting of random noise caused by birefringence.

Above methods  have been successfully applied in  OVS system, but they still have some limitations. The dual-path and dual-crystal methods work well in constant temperature environments but cannot withstand rigorous temperature cycling experiments, which could alter the optical path and the distribution of stress birefringence, leading to compensation failure. The image demodulation method requires expensive optical elements, and  its relatively complex structure is susceptible to external vibration.

In this paper, we propose a method to regulate and suppress birefringence based on arbitrary electric field direction modulation (AEFDM) and isomerism electrodes. Through thermal, mechanical and optical multi-physics  Finite Element Method (FEM) simulation, it is found that the metal or transparent isomerism electrodes could regulate the  thermal stress distribution in optical path, and the induced birefringence and measurement error could be depressed by calculation with the aid of Jones Matrix and photoelastic theory.  The experiment is conducted in 10 mm cubic BGO crystal with isomerism copper or Indium Tin Oxide (ITO) electrodes. The results demonstrated that the Cu isomerism electrodes with electric field angle of 59.9 degrees could generate 37\% less birefringence error, compared to standard transverse modulation. While the ITO isomerism electrodes could produced 7-times larger birefringence error due to its bad ductility and thermal conduction.

\section{\label{sec:level1} Analysis  of error induced by thermal stress birefringence}

\subsection{\label{sec:level2}Simulation of thermal stress in OVS header}

When an electro-optic crystal is subjected to a variable temperature environment or non-uniform temperature field, thermal expansion will cause the electro-optical crystal to undergo certain deformation. The expression of induced stress$^{12-14}$ are shown in Equations (1)(2),

\begin{eqnarray}
	\sigma_{ij} &=& \frac{E}{1-\nu} [\epsilon_{ij}+\frac{\nu}{1-2\nu}\mathbf{Tr}[\bar{\epsilon}]\delta_{ij}-\frac{1+\nu}{1-2\nu}\alpha T \delta_{ij}] \\
	\epsilon_{ij} &=& \frac{1}{2}[\partial U_i/\partial x_j + \partial U_j/\partial x_i]
\end{eqnarray}
, where $E$ is Young's modulus, $\alpha$ and $\nu$ is the material's thermal coefficient and Poisson's ratio, $\mathbf{Tr}[\bar{\epsilon}]$ is the  trace of strain matrix $\epsilon$  ,  $\delta_{ij}$ is the Kronecker delta ($i$,$j$ = 1,2,3), $\epsilon_{ij}$ is the strain tensor element, \textbf{U} is the volume element displacement, $x$ is the spatial coordinates and $T$ represents temperature.

In an equilibrium state, the internal stresses within any volume unit should be balanced. Assuming that there are no internal stresses within the crystal, the equilibrium condition is  $\partial{ \sigma_{ij} } / \partial x_j = 0$ where $\sigma_{ij}$ is the stress component. The volume element displacement vector \textbf{U} satisfy 
\begin{eqnarray}
	\nabla^2 \mathbf{U} + \frac{1}{1-\nu} \nabla(\nabla \cdot \mathbf{U}) = \frac{2(1+\nu)}{1-2\nu}\alpha T 
\end{eqnarray}
with boundary condition $\sigma_{ij} n_j = 0$, and $n_j$ is the normal vector in direction $x_j$.

According to Equations (1-3), the FEM simulation is conducted, and its model is shown in FIG. 1. The top cube is the BGO crystal and has a size of 10 mm * 10 mm * 10 mm; The middle cylinder is made of silica and served as a heat conductor, its radius and height are 25 mm and 15 mm respectively; The bottom cuboid represents the aluminum (Al) heating platform with a size of 50 mm * 50 mm * 10 mm.  
The copper or ITO electrodes are directly bonded to the BGO crystal, and the crystal is placed on the cylinder silica with hard connection. The silica cylinder is also served to keep certain space between the BGO and metal heating platform, since the latter would disturb the electric field distribution inside BGO. 

The key simulation condition and parameters are set as following. The transient mode is selected, and the total simulation time is 60 s with step time of 5 s. Convective heat flux mode is adopted as the heat flux condition, in order to emulate the heat exchange process between the BGO crystal and  surrounding flowing air. The temperature of environment is set to 300 K, and the bottom heater's temperature is configured to 358 K. The related parameters of different materials in system are listed in Table I.

\begin{figure}[t]
	\includegraphics[width=0.35\textwidth]{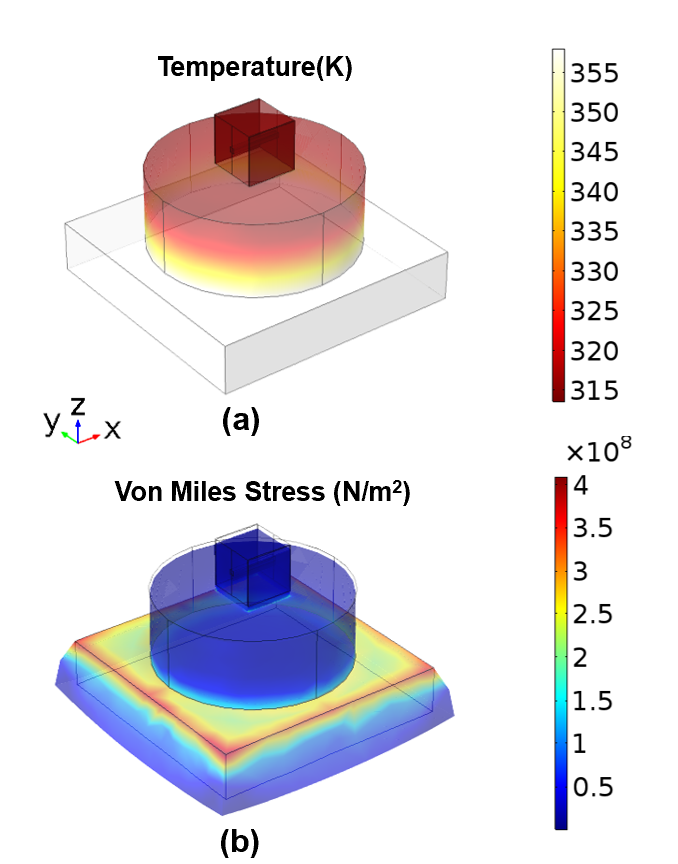}
	\caption{\label{fig:epsart}
		FEM simulation model and results. (a) Simulated temperature distribution. (b) Simulated model's stress.  
	}
\end{figure}

\begin{table}[pbp]
	\caption{\label{tab:table4}FEM Simulation parameters in OVS model  }
	\begin{ruledtabular}
		\begin{tabular}{ccccc}
			Parameter &BGO&SiO$_2$&Cu(elecrodes) & Al(plate)\\
			\hline
			$\rho(\mathrm{g} \cdot \mathrm{cm}^{-3})$&7.13&2.20&8.94&2.73\\
			$c(\mathrm{J} \cdot \mathrm{kg}^{-1} \mathrm{K}^{-1} )$ &3.5e2& 8.91e2 & 3.85e2 & 7.50e2 \\
			$\nu$
			&0.20
			& 0.17 & 0.34 & 0.33 \\
			$E (\mathrm{Pa})$ 
			&7.31e10 &7.50e10 &12.6e10 &7.0e10 \\
			$\alpha (\mathrm{K}^{-1})$ 
			&6.3e-6 &5.5e-7 &1.7e-5 & 2.4e-5\\ 
			$K(\mathrm{W} \cdot \mathrm{m}^{-1}  \mathrm{k}^{-1}))$ 
			&0.18 &1.46 &400 &238
		\end{tabular}
	\end{ruledtabular}
\end{table}

FIG. 1(a) shows the temperature distribution of the model for t = 60 s, where the standard transverse modulation and copper electrodes are applied. The bottom heater's temperature is kept in 85 degree Celsius, while The BGO's temperature could be heated to 42 degrees because the model is in a opened air environment.

FIG. 1(b) plots the distribution of system's Von Mises stress, which is a popular parameter to express the material's inner stress in engineering application. Its formula is give by 
\begin{eqnarray}
	\sigma_{von}=& &\frac{1}{\sqrt{2}} [(\sigma_{11}-\sigma_{22})^2+(\sigma_{22}-\sigma_{33})^2+(\sigma_{33}-\sigma_{11})^2 \nonumber \\
	& &+6(\sigma_{12}^2+\sigma_{23}^2+\sigma_{31}^2)]^{1/2} 
\end{eqnarray}
, where $\sigma_{11}$, $\sigma_{22}$ and $\sigma_{33}$ are the three main stress components, and $\sigma_{12}$, $\sigma_{23}$, $\sigma_{13}$ are the tangential stress components. To display the material's stress and strain clearly, the deformation of the geometry has been over-exaggerated.  The six components of stress along optical path are shown in FIG. 2, and the three main stress and one tangential components $\sigma_{13}$ are in the dominance.

\begin{figure}[htpb]
	\includegraphics[width=0.3\textwidth]{ 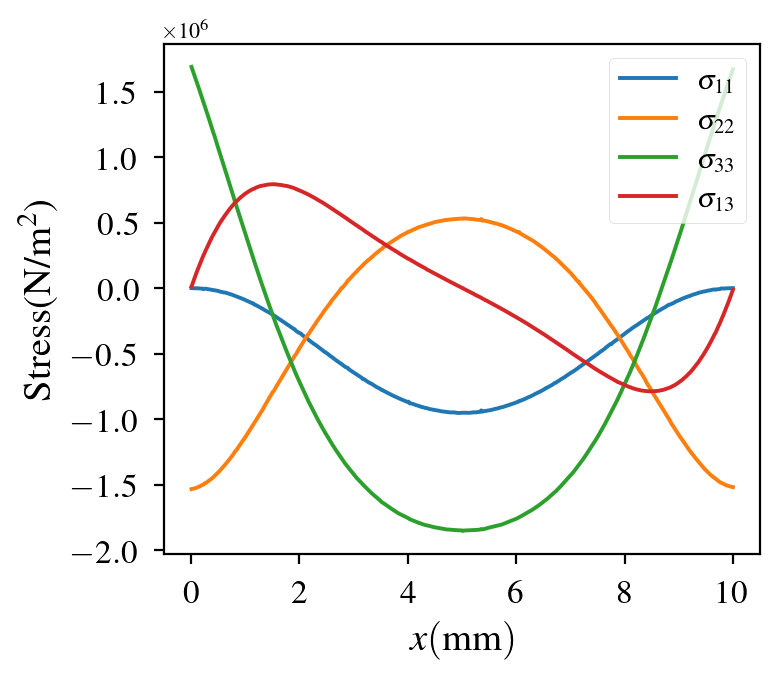}
	\includegraphics[width=0.3\textwidth]{ 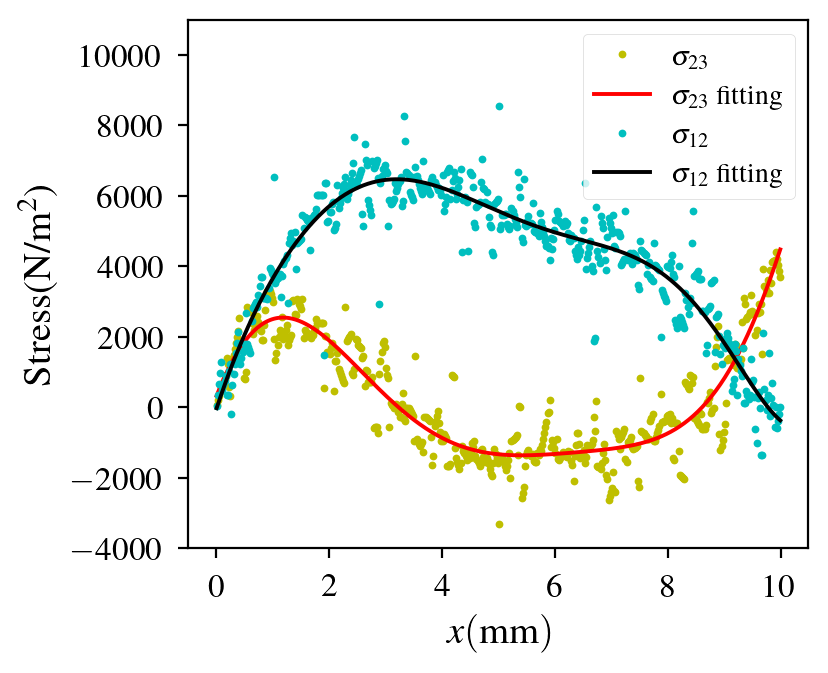}
	\caption{
		Stress distribution in BGO crystal along optical path. The top sub-graph shows the distribution of  $\sigma_{11}$, $\sigma_{22}$ and $\sigma_{33}$ and $\sigma_{13}$. The bottom one shows that of $\sigma_{12}$, $\sigma_{23}$, which are much smaller than the other 4 components. } 
	
\end{figure}

\subsection{\label{sec:level2}Calculation methods of BGO's Birefringence  }

Acorrding to Ellipsoidal Theory of refractive index$^{24}$, any crystal's refractive could be described by
\begin{eqnarray}
	\sum_{i,j=1}^3B_{ij}x_{ij}=1
\end{eqnarray}
, where $B_{ij} = 1/n_{ij}^2$ and $n_{ij}$ are the refraction. Since BGO is an isotropic crystal, its ellipsoidal of refractive index is a sphere for arbitrary coordinate, and the cross term of $B_{ij}$ equal to 0.

Thermal stress makes the refractive ellipsoidal change, following the rule of photo-elastic effect described by$^{24}$ 
\begin{eqnarray}
	\begin{bmatrix}
		\Delta B_{11} \\ \Delta B_{22} \\ \Delta B_{33} \\ B_{12}
		\\B_{23} \\B_{31}   
	\end{bmatrix}
	=\mathbf{Q} \begin{bmatrix}
		\sigma_{11} \\ \sigma_{22} \\ \sigma_{33} \\ \sigma_{12}
		\\\sigma_{23} \\ \sigma_{31}  
	\end{bmatrix}
\end{eqnarray}
, where \textbf{Q} is the photo-elastic matrix in the simulation coordinate and its expression is given in Appendix A.

For light travel in the $x_{ii}$ direction and $i=1,2,3$, the refraction of fast light and slow light are$^{13,14}$
\begin{eqnarray}
	n_{\pm} = \bigg[\frac{1}{2} ( B_{jj}+B_{kk} \pm \sqrt{(B_{jj}-B_{kk})^2+4 B_{jk}^2}  )\bigg]^{\frac{1}{2}}
\end{eqnarray} 
Note that $|B_{jj} - B_{0} |<<B_{0}$, $B_{0} = 1/n_o^2$ and $n_o$ are the original refractive index of BGO, the difference of refraction could be simplified as
\begin{eqnarray}
	|\Delta n| = \frac{1}{2}n_o^3 \sqrt{(B_{jj}-B_{kk})^2 + 4B_{jk}^2}
\end{eqnarray} 
, while the roation angle of the fast or slow axis is 
\begin{eqnarray}
	\theta = \frac{1}{2} \arctan\bigg|\frac{2B_{jk}}{B_{jj}-{B_{kk}}}\bigg|.
\end{eqnarray} 

 To calculate stress induced birefringence, namely $|\Delta n|$ and $\theta$, we should know	$(B_{jj}-B_{kk})$ and $B_{jk}$  according to Equations (8)(9). In our simulation or experiment, the light is travel through $x_{11}$ direction, thus $i=1$, $j=2$, $k=3$. With the help of Equations (6)(A4), we get
\begin{eqnarray}
	B_{22}-B_{33} =& & \frac{1}{2}(q_{11}-q_{12})(\sigma_{11}+\sigma_{22}-2\sigma_{33})\nonumber\\& &- \frac{1}{2}q_{44}(\sigma_{11}-\sigma_{22}) ,\\
	B_{23} = & & q_{44}\sigma_{23}
\end{eqnarray} 
, where photo-eleatic coefficients of BGO satisfy$^{25}$ $q_{11}-q_{22}$ = -2.995E-13 $\text{m}^2$/N, and $q_{44}$ = -1.365E-12 m$^2$/N .

\subsection{\label{sec:level2} Analysis of accumulation error along optical path}

FIG. 3 shows the set-up of the OVS based on BGO crystal in standard transverse modulation, which is composed of polarizer, BGO crystal, quarter wave plate (QWP), analyzer and photo diode. Since the stress is different everywhere in optical path, the birefringence and its induced error should be calculated in every tiny section, and then accumulated by using Jones Matrix. The calculation process is described as follows. 

\begin{figure}[bp]
	\includegraphics[width=0.35\textwidth]{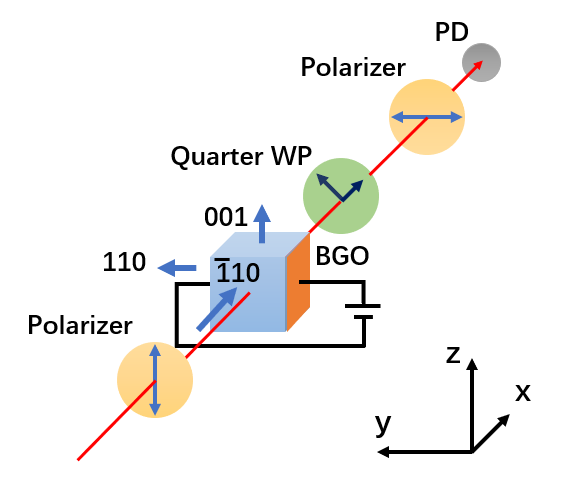}
	\caption{\label{fig:epsart} 
		OVS's set-up for transverse modulation.}
\end{figure}

The normalized Jones Vector of the input light after the polarizer is 
\begin{eqnarray}
	\mathbf{E}_{in} = \begin{bmatrix}
		1  \\
		0
	\end{bmatrix}
\end{eqnarray}

The QWP and analyzer’s Jone Matrixes, \textbf{Q} and \textbf{L}, are given by
\begin{eqnarray}
	\mathbf{Q} &=& \frac{1}{\sqrt{2}}\begin{bmatrix}
		1 & i \\
		i & 1
	\end{bmatrix}\\
	\mathbf{L} &=&\begin{bmatrix}
		0 & 0 \\
		0 & 1
	\end{bmatrix}.
\end{eqnarray}

The optical path is divided into $m$ sections, where the stress could be considered as homogeneous. The Jones Matrix of each section is 
\begin{eqnarray}
	\mathbf{J}_{m} = \mathbf{R}_m^{-1} \mathbf{P}_{m} \mathbf{R}_m
\end{eqnarray}
, where $\mathbf{P}_m$ and $\mathbf{R}_m$ are the phase delay matrix and rotation matrix in the $m$th section. Their relationship to birefringence is given by

\begin{eqnarray}
	\mathbf{P}_m &=& \begin{bmatrix}
		\exp (i \delta_m/2) & 0 \\
		0 & \exp (-i \delta_m/2)
	\end{bmatrix}\\
	\mathbf{R}_m &=& \begin{bmatrix}
		\cos \theta_m & \sin \theta_m \\
		-\sin \theta_m & \cos \theta_m
	\end{bmatrix} \\
	\delta_m &=& \frac{2\pi}{\lambda}L_m \Delta n_m
\end{eqnarray}
, where $\delta_m$ is the section's phase delay and $L_m$ is the length of the section, $\lambda$ is the wavelength, $\delta_m$ and $\theta_m$ are computed with Equations (8-11).

The output light's electric vector and intensity are    
\begin{eqnarray}
	\mathbf{E}_{out} &=&\mathbf{L} \mathbf{Q} \bigg( \prod \limits_{m=0}\mathbf{J}_m \bigg)	\mathbf{E}_{in}\\ 
	I_{out} &=& \mathbf{E}_{out} \mathbf{E}_{out}^*.
\end{eqnarray}

The birefringence induced error is the difference between the output light intensity and the static work point. Since the input light power is normalized to 1, the static work point of OVS could be calculated from intensity equation of OVS, 
\begin{eqnarray}
	I= \sin^2[\frac{1}{2}( \frac{V}{V_{\lambda/2}}+\frac{\pi}{2})]
\end{eqnarray}
, where $V_{\lambda/2}$ is the half-wave voltage (HWV). By setting the applied voltage $V$ = 0, the static point equal to 0.5, and the error is given by
\begin{eqnarray}
	Error = I_{out} - 0.5.
\end{eqnarray}

\section{Birefringence Regulation method based on AEFDM}

\begin{figure}[pbb]
	\centerline{\includegraphics[width=0.35\textwidth]{ 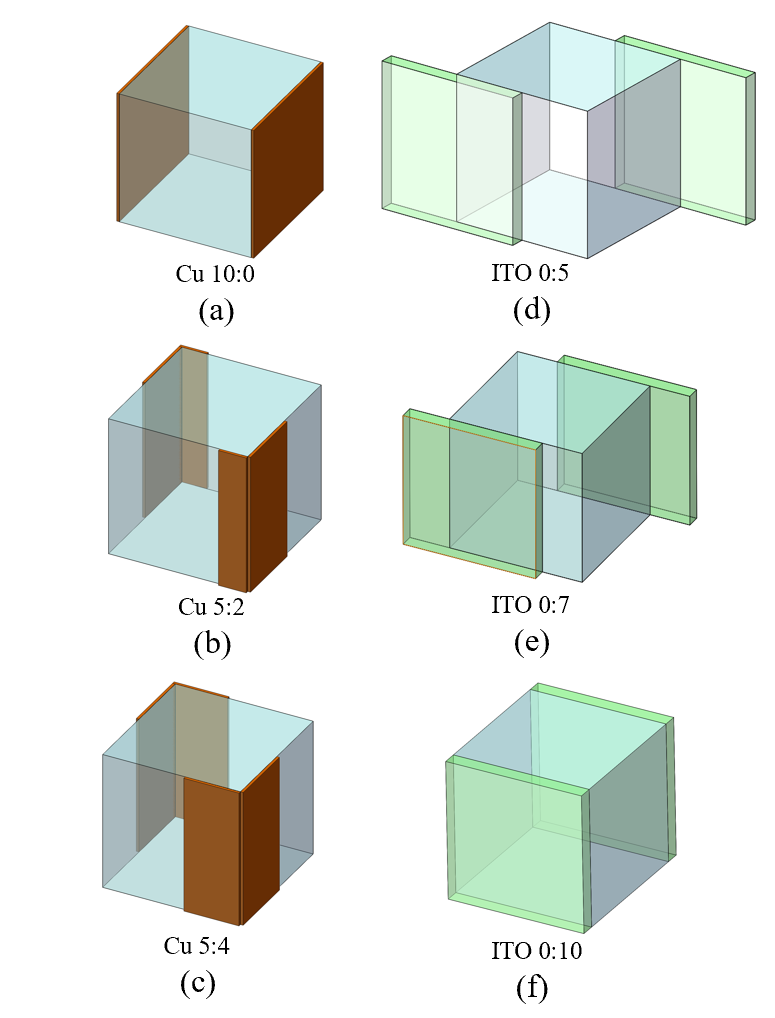}}
	
	\caption{
		Schematic of 6 different modulation mode, named by the material of electrodes and the ratio of electrode sizes in x  and y directions. (a) Cu 10:0, transverse modulation, (b) Cu 5:2, (c) Cu 5:4, (d) ITO 0:5, (e) ITO 0:7, and (f) ITO 0:10.}
	
	\label{fig3}
\end{figure}

The AEFDM mode is a special modulation mode where angle between light and electric field could be arbitrary value from 0 to 180 degrees$^{26-28}$. To regulate the electric field direction, the isomerism electrodes made of copper coil or ITO are adopted. 

FIG. 4 shows the transverse modulation configuration and 5 different AEFDM mode's set-up. For angles whose values falls within the range from 90 degrees and 45 degrees, L-shaped electrodes are proven to produce a more uniform field in the region near the optical path compared to staggered ones$^{26}$.
In the case where the angle is between 45 and 0 degrees, ITO$^{29}$ transparent electrodes are employed because the metal ones would block the input polarized light.  The 6 modulation modes were named by their material and the ratio of  electrode length in x and y direction, and  called Cu 10:0, Cu 5:2, Cu 5:4, ITO 0:5, ITO 0:7 and ITO 0:10.
Their average electric field direction in optical path and HWV are listed in Table II. The HWV becomes larger as the field direction angle with respect to light direction decreases.

\begin{table}[ptp]
	\caption{\label{tab:table4}Electric field angle and HWV of different modulation modes.}
	\begin{ruledtabular}
		\begin{tabular}{rcl}
			
			Modulation Mode &
			Field Angle (degrees) &
			
			HWV (kV) \\		
			\hline		
			Cu 10:0& 89.99 &    47.06 \\
			Cu 5:2 & 71.79 &   63.57 \\
			Cu 5:4& 59.87 &    52.53 \\
			ITO 0:5& 29.61 &   74.04 \\
			ITO 0:7& 7.63 &   341.88 \\
			ITO 0:10&  0.00 & 1409.45 \\		
			
		\end{tabular}
	\end{ruledtabular}
\end{table}

The transient simulation of temperature and stress has been conducted on 6 different modes. The Von Mises stress distribution in the crystal’s cross section at time stamp of 60 s are shown in FIG. 5. Due to the edge effect, stress in 4 AEFDM modes (FIG. 5(b-e)) is more concentrated at the contact point between the electrode edge and the crystal, thus affecting the stress distribution along the linear polarized light's path  labeled by black arrows.

\begin{figure}[htbp]
	\includegraphics[width=0.42\textwidth]{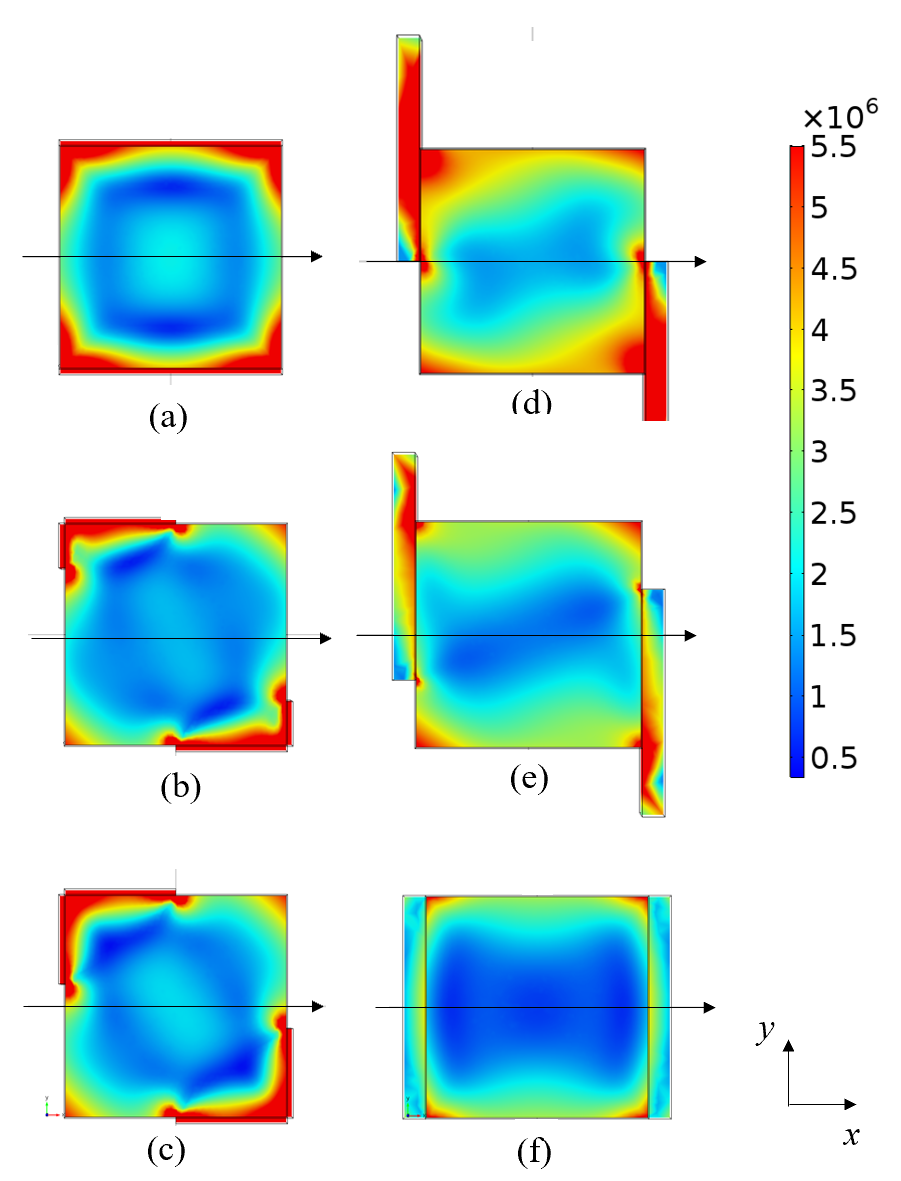}
	\caption{\label{fig:wide}
		Von Miles stress in x-y cross-section for different modulation mode.(a) Cu 10:0, transverse modulation, (b) Cu 5:2, (c) Cu 5:4, (d) ITO 0:5, (e) ITO 0:7 and (f) ITO 0:10.}
\end{figure}

With the aid of Equations (8-22), the accumulated birefringence error at different crystal temperature are computed for 6 modulation modes. The theory calculation results are shown in FIG. 6. 

FIG. 6 (a) presents the calculated error in the standard transverse modulation with copper electrodes. At an initial temperature of 27°C, the stress-induced birefringence error is 0. Upon start heating the bottom stage, stress is instantaneously generated within the crystal, causing a step change in measurement error. Then, the error increases slowly and gradually stabilizes. Within the temperature variation from 27°C to 42°C, the total birefringence error is 4.2E-4.

FIG. 6 (b) shows the error of Cu 5:2 mode with field angle of 71.2 degrees. The step phenomenon in initial temperature point can also be observed, but its magnitude of change is significantly smaller. The error then changes approximately linearly with temperature, and the total induced error in temperature window is 2.8E-4, which is 70\% of the error in transverse modulation.

FIG. 6 (c), corresponding to mode Cu 5:4, displays a negative step error at the initial temperature, and thereafter, the error changes positively in an exponential function curve and stabilizes around 42°C. The absolute value of the changed error is less than 2.1E-4, which is smallest. It should be noted  in real experiment, the heating stage temperature raised smoothly, and no instantaneous step change of error was observed.

\begin{figure}[tbhp]
	\includegraphics[width=0.47\textwidth]{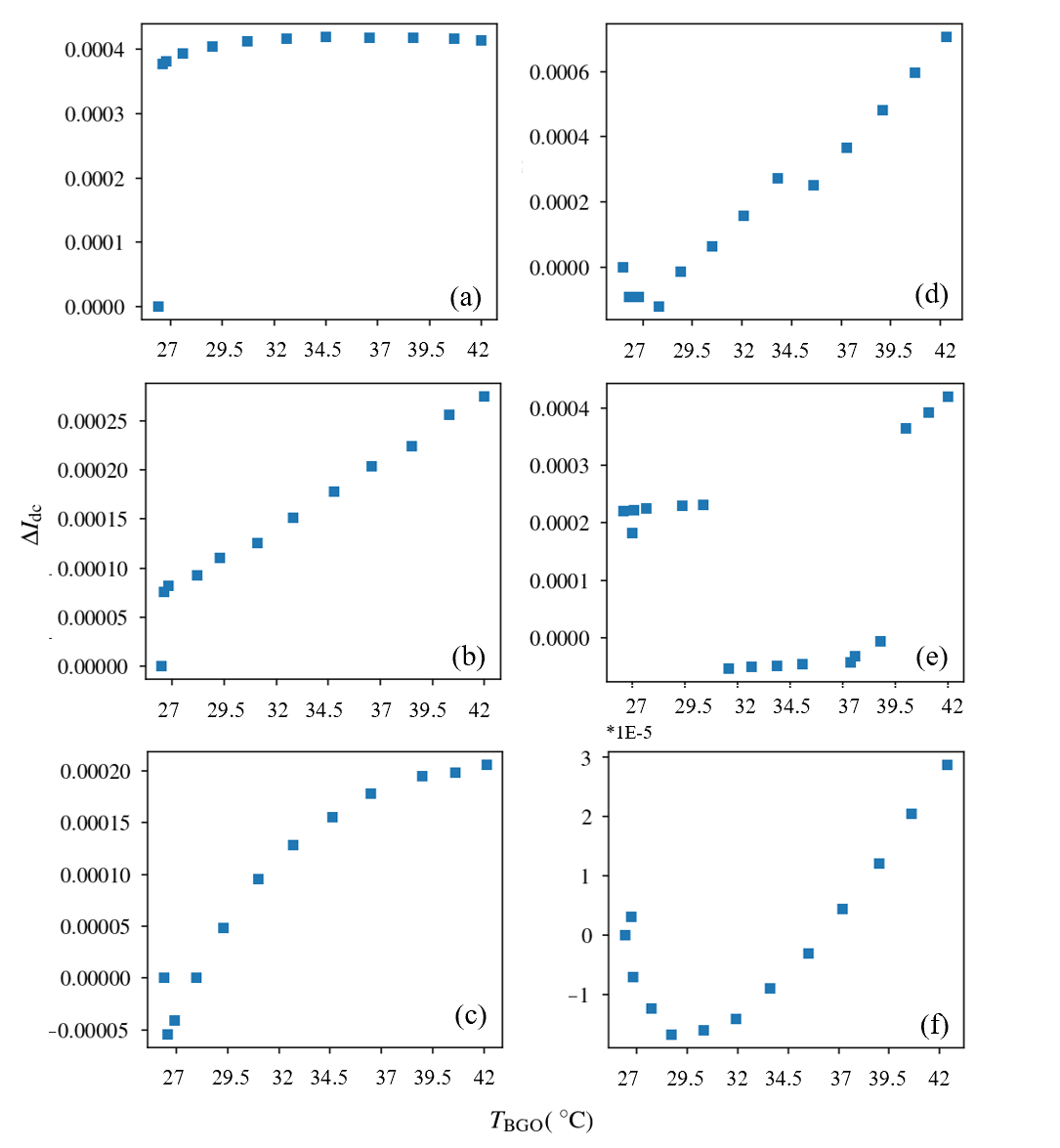}
	\caption{\label{fig:epsart}Calculated birefringence error and its relationship with temperature of BGO for 6 mode: (a) Cu 10:0, transverse modulation, (b) Cu 5:2, (c) Cu 5:4, (d) ITO 0:5, (e) ITO 0:7 and (f) ITO 0:10.}
\end{figure}

FIG. 6(d)(e)(f)  shows the calculation error for three modes using ITO transparent electrodes, with the electric field angles generated  equal to 26.61, 7.63 and 0 degrees, respectively. As the crystal temperature increases, the stress birefringence error first increases and then decreases. As the angle of the electric field becomes smaller, the error variation within the temperature range will also reduce.

\section{Experiment}

The experimental setup was constructed on an optical platform, as depicted in FIG. 7(a). The light source employed was a 976nm semiconductor distributed feedback (DFB) laser with polarization-maintaining fiber output. Gran-Taylor prisms were used as the polarizer and analyzer. A silicon optical power meter was utilized to measure the output light power, and its output analog signals were sampled by a 16-bit acquisition card (model NI USB-6361) at a sampling rate of 100 kHz. The 10 mm cubic BGO crystal used was cut along the directions  
$\bar{1}10$, 110, and 001. The copper electrodes were made of 0.5mm copper foil, while the ITO electrode's size was 10 mm * 10 mm * 1 mm. These electrodes were adhered to the BGO by using ultraviolet curing adhesive. The metal wire, coated with an insulation layer, was connected to the electrode through welding or silver glue. A type K thermocouple was attached directly above the BGO, as shown in the inset of FIG. 7(a). The BGO crystal was placed on a cylindrical silica support and heated by a metal heating platform.

\begin{figure}[bp]
	\includegraphics[width=0.35\textwidth]{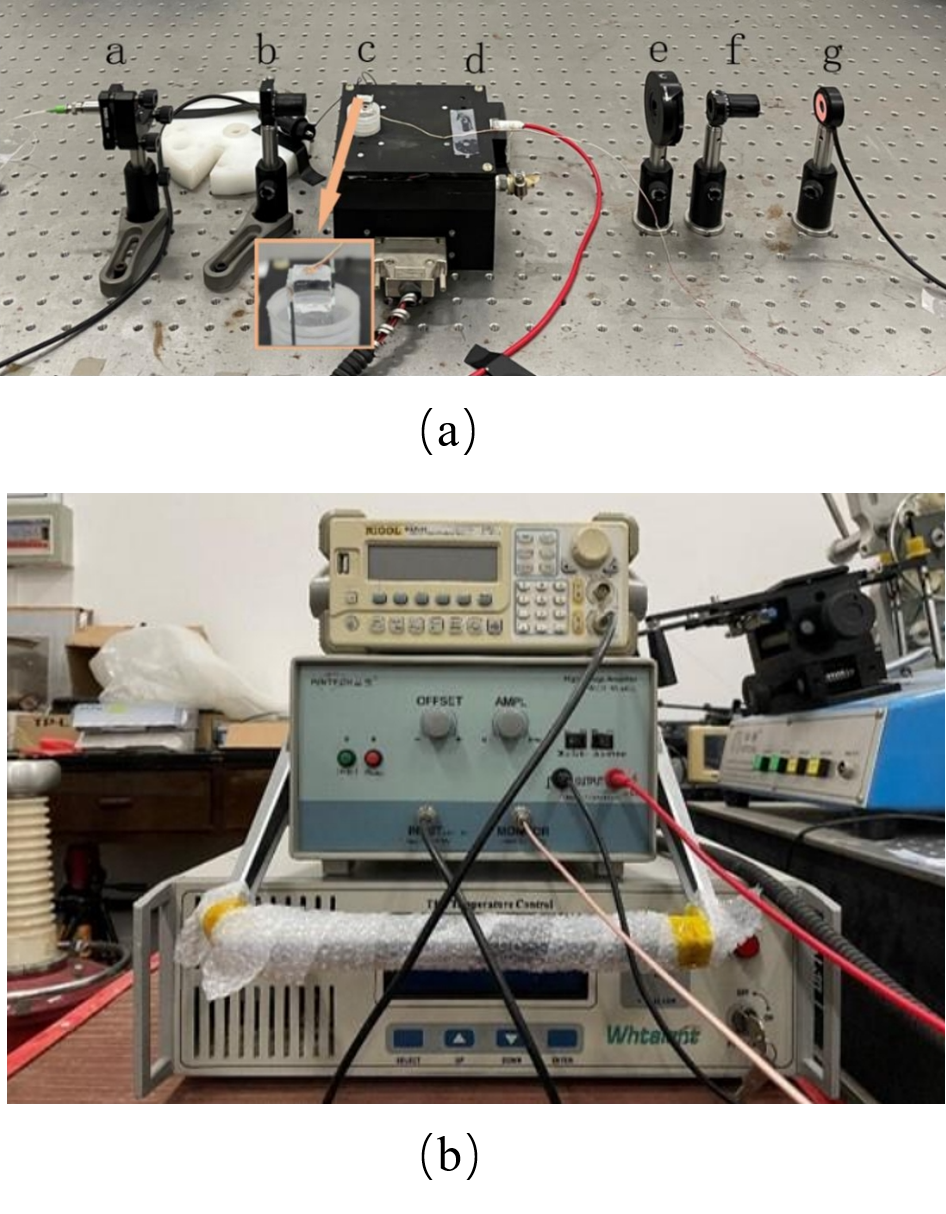}
	\caption{\label{fig:epsart}Experimental set-up.(a) Optical path of OVS, where a is the DFB laser's collimator on a mount, b is the Gran-Tayer prism polarizer, c is the BGO crystal with electrodes on a silica cylinder, whose expanded images is shown in the inset, d is the heating platform, e is the QWP, f is the Gran-Tayer analyzer, and g is a silicon light detector. (b) Electronics for generating high voltage and controlling the temperature. The top one is a signal generator, the middle is a high-voltage amplifier, and the bottom is a PID temperature controller. }
\end{figure}

\begin{figure}[tbh]
	\includegraphics[width=0.45\textwidth]{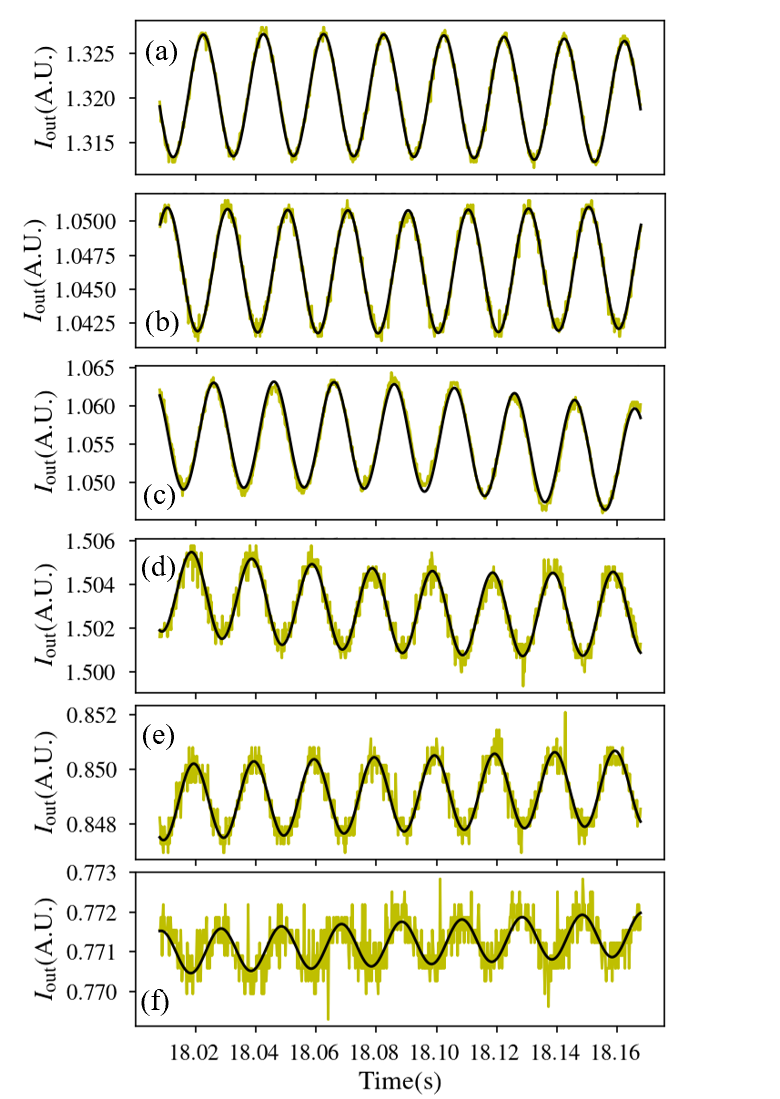}
	\caption{\label{fig:epsart} Output intensity of OVS with input 50Hz AC voltage for different modes. (a) Cu 10:0, (b) Cu 5:2, (c) Cu 5:4, (d) ITO 0:5, (e) ITO 0:7 and (f) ITO 0:10.}
\end{figure}

\begin{figure}[htpb]
	\includegraphics[width=0.46\textwidth]{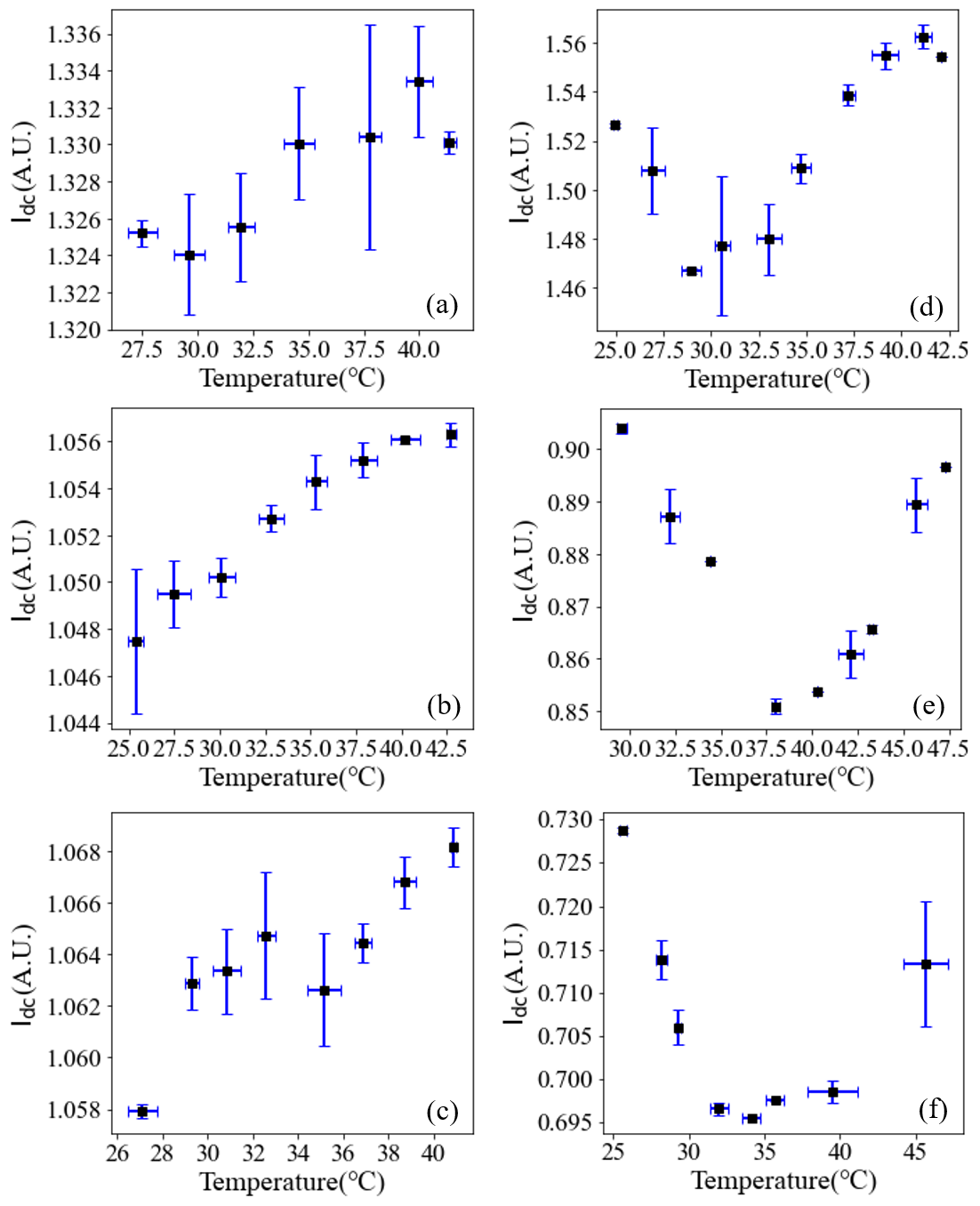}
	\caption{\label{fig:epsart} Experimental results of stress birefringence error with temperature variation under six different modes: (a) Cu 10:0, (b) Cu 5:2, (c) Cu 5:4, (d) ITO 0:5, (e) ITO 0:7 and (f) ITO 0:10.}
\end{figure}

FIG. 7(b) shows the utilized electronic devices. The top and middle devices are a signal generator and a high-voltage amplifier, respectively. They were used to generate the AC high voltages applied to the electrodes. The bottom device is a Proportional-Integral-Derivative (PID) temperature controller, used for regulating the temperature of the heating platform. At the start of the experiment, the target temperature of the heater was set directly to 85 degrees Celsius. The BGO crystal with different types of electrodes was heated to approximately 45 degrees Celsius in the end, due to the open environment. The  heating time  to ensure stabilization of the BGO's temperature was approximately 55 minutes. The signals from the electrode voltage, photodiode and thermocouple were simultaneously recorded by the acquisition card.

FIG. 8 recorded the OVS signals of six modes,  with a total duration of 160 ms and 8 complete periods. The birefringence error in measured signals has a main frequency much lower than the power frequency (50 or 60 Hz). It can be regarded as the drift of a DC signal in a short time range such as about 100 ms, and unrelated to the measured AC voltage. To extract the birefringence error, the measured signals in FIG.8 are fitted by 
\begin{eqnarray}
	I_{out} =&& I_{\text{AC}} \cos{(2\pi*50t + \phi)} + I_{\text{DC}} \\
	I_{\text{DC}} =&& at^2+bt+c
\end{eqnarray}
, where $I_{\text{DC}}$ means the DC drift of signals induced by birefringence, $t$ is time and $\phi$ is the  original signal phase. The DC signals are further fitted by a quadratic function, in which \textit{a}, \textit{b} and \textit{c} are the fitting coefficients.
For signal statistics and analysis, the stress birefringence error is taken as the average value of the fitted $I_{DC}$, and the temperature is considered as the average temperature over this time window.

FIG. 9 shows the experimental results of stress birefringence error with respect to temperature under six different modes. When OVS adopts modulation modes based on copper electrodes, the stress error increases nearly monotonically with temperature. The Cu 5:2 and  Cu 5:4 electrodes has less birefringence error in the temperature window compared to Cu 10:0 (transverse modulation). For the three modulation mode with ITO transparent electrodes, the birefringence error decreases first and then increases with temperature, and the trend of change is non monotonic. Their total drift error resulted from birefringence is larger than that with the copper electrodes.

\begin{figure}[hpbt]
	\includegraphics[width=0.38\textwidth]{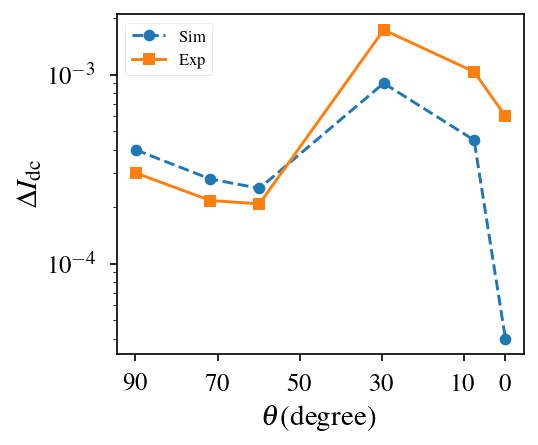}
	\caption{\label{fig:epsart} Relationship between the total birefringence  error and the electric field direction of modulation mode. The orange squares and solid lines represent the experimental results, and the blue circles and dashed lines represent the simulated data. }
\end{figure}

The experimental and simulation comparison results is shown in FIG. 10. The total stress birefringence error in the temperature windows in the experiment, as the change of electric field angle, has the similar trend with that in the simulation. This indicates the effectiveness of  stress birefringence simulation method proposed in this paper.
Since the time scales of simulation and experiment are different, with the former being 1 minute and the latter averaging 55 minutes. In order to better compare experimental data and simulation results, we adopted the data processing method of fiber optic gyroscope$^{30,31}$ to eliminate the influence of time on random drift error. The detail data process method has been discussed in Appendix B.

The above results also lead to the following conclusions: For two modulation cases with copper or ITO electrodes, the total birefringence error reduce as the angle of the electric field decreases, respectively.   The utilize of AEFDM and copper isomerism electrodes Cu 5:4 (with electric field angle 59.9 degrees) can effectively reduce 37\% stress birefringence error, compared to the transverse modulation mode (Cu 10:0 with field angle of 90 degrees). However, both simulation calculations and experimental results indicate that  ITO transparent electrodes produce more error than the transverse modulation. For the worst case, the ITO electrodes with field angle of 29.6$^{\circ}$ (ITO 0:5) produces  approximately 7 times error as the transverse modulation does. This may attributed to the poor ductility and thermal conductivity of the ITO electrode's fused silica substrate.

\section{Conclusion}
This paper proposes a stress birefringence error suppression method for OVS based on isomerism electrodes and the AEFDM. Using a combined simulation approach based on FEM, Photoelastic Theory, and the Jones Matrix, we found that isomerism electrodes made of copper foil or ITO in the AEFDM mode could regulate the BGO's stress in optical path and its induced birefringence error. Experiments verify the simulation results, and demonstrate that  copper isomerism electrodes with electric field angel of  59.87 degrees could reduced the error by 37\%, within the temperature range of 25°C to 40°C. However, the use of ITO isomerism electrodes with a fused silica substrate fails to reduce the stress birefringence error, potentially due to the poor thermal conductivity and ductility of the substrate. The analysis method of stress birefringence error proposed in this paper has proven to be effective through experiments, and can be widely applied to electro-optic crystals with arbitrary field distribution. The proposed stress birefringence error suppression method can also be broadly utilized in the design of high-precision OVS and electro-optic modulators.

\begin{acknowledgments}
	This work is supported by the National Natural Science Foundation of 
	China (No.51807030 and No.51977038), the Fujian Provincial Science and Technology 
	Department guiding program (No.2017H0013). 
\end{acknowledgments}

\section*{Data Availability Statement}

The data that support the findings of this study are available from the corresponding author upon reasonable request.

\begin{appendix}

\appendix

\section{Photo-elastic Matrix in Simulation Coordinate}

In our simulation and experiments, the light propagates along x direction, correspond to crystal direction $\overline{1}10$, and the electric field is applied in y drection, the crystal 110 direction. Thus the simulation coordiate and the crystal coordate has a transformation relationship, which is described by 

\begin{eqnarray}
	\mathbf{T} = \begin{bmatrix}
		-\frac{\sqrt{2}}{2} &\frac{\sqrt{2}}{2} & 0\\
		\frac{\sqrt{2}}{2} &\frac{\sqrt{2}}{2}  & 0 \\
		0 & 0 & 1
	\end{bmatrix}.
\end{eqnarray} 

By using tensor rotation equation $\mathbf{T} ^{-1} \mathbf{ \sigma} \mathbf{T} $,  the transform matrix $\mathbf{A}$ related to simplified 6 components follows 
\begin{eqnarray}
	\sigma' =\mathbf{T}^{-1} \begin{bmatrix}
		\sigma_{11} & \sigma_{12} & \sigma_{13}\\
		\sigma_{12} &\sigma_{22} & \sigma_{23} \\
		\sigma_{13} & \sigma_{23} & \sigma_{33}
	\end{bmatrix} \mathbf{T} 
	= \mathbf{A} \begin{bmatrix} 
		\sigma_{11} \\ \sigma_{22} \\ \sigma_{33} \\\sigma_{12}
		\\ \sigma_{23} \\ \sigma_{13}
	\end{bmatrix}.
\end{eqnarray} 
The expression of $\mathbf{A}$ is then
\begin{eqnarray}
	\mathbf{A} = \begin{bmatrix}
		\frac{1}{2} & \frac{1}{2} & 0 & 0 &0 & -\frac{1}{2} \\
		\frac{1}{2} & \frac{1}{2} & 0 & 0 &0 & \frac{1}{2} \\
		0 & 0 & 1 & 0 &0 & 0 \\
		0 & 0 & 0 & \frac{\sqrt{2}}{2} &\frac{\sqrt{2}}{2} & 0 \\
		0 & 0 & 0 &\frac{\sqrt{2}}{2} &-\frac{\sqrt{2}}{2} & 0 \\
		-\frac{1}{2} & \frac{1}{2} & 0 & 0 &0 & 0 \\		
	\end{bmatrix}.
\end{eqnarray} 

The photo-elastic matrix for BGO crystal in crystal coordinate  has only 3 independent components $q_{11}$, $q_{12}$, $q_{44}$ and  expression of the matrix is
\begin{eqnarray}
	\mathbf{Q}^{CRY} = \begin{bmatrix}
		q_{11} & q_{12} & q_{12} & 0 & 0 & 0\\
		q_{12} & q_{11} & q_{12} & 0 & 0 & 0 \\
		q_{12} & q_{12} & q_{11} & 0 & 0 & 0 \\
		0 & 0 & 0 & q_{44}  & 0 & 0 \\
		0 & 0 & 0 &0  & q_{44} & 0 \\
		0 & 0 & 0 & 0  & 0 & q_{44}
	\end{bmatrix}.
\end{eqnarray}
Thus using the tensor's coordinate transformation equation, the photo-elastic matrix in simulation coordinate is given by 
\begin{eqnarray}
	\mathbf{Q} = \mathbf{A} \mathbf{Q^{CRY}} \mathbf{A}^{-1}. 
\end{eqnarray}

\section{Method to Eliminate the influence of time on Drift Error in OVS}

The DC component drift extracted from Formulas (23) and (24) is defined as the bias instability of OVS, and the power spectral density of this type of random disturbance signal $^{30,31}$ is expressed as:

\begin{eqnarray}
	S_{\Omega, BI}=\left( \frac{B_s^2}{2\pi} \right) \frac{1}{f}
\end{eqnarray}
, where $B_s$ is the component amplitude of noise, and \textit{f} is the component's frequency.   

The time-domain signal of zero bias instability can be estimated by Equation (B2)$^{30}$, which could be interpreted as 
the mean square value of the zero point random disturbance
\begin{eqnarray}
	\sigma_{\Omega,BI} = \sqrt{S_{\Omega, BI}\tau}
\end{eqnarray}
, where $\tau$ is the total measurement time.

Suppose the low frequency zero-drift error in the heating process  has the period compatible to the total measurement time. To simplify the analysis, we follow the method in Ref. 30 and set $f=1/ \tau$. By substituting Formula (B1) into (B2), the bias instability equal to

\begin{eqnarray}
	\sigma_{\Omega,BI} =  \frac{B_s^2}{2\pi} \tau
\end{eqnarray}
. Equation (B3) indicates that the bias will increase linearly as the total measurement time. 

In order to compare the experimental measurement results with the simulation, Equation (B3) is adopted to correct the influence of time to random birefringence error.
The amplitude of noise in frequency domain $B_s$ in simulation and experimental system could be considered as the same, since the parameters and conditions used in the simulation are nearly consistent with the experimental results as shown in FIG. 1(a) and FIG. 7(a). It is necessary to correct the simulation or experiment results by removing the time coefficient $\tau$. Following this process, Fig. 10 give the comparison between the real measurement results in FIG. 9 and simulation results in FIG. 6.

\nocite{*}
\end{appendix}

\section*{References}
\bibliography{JAP_REF}

\end{document}